\documentclass[pra,twocolumn,showpacs]{revtex4}

\usepackage[english]{babel}
\usepackage[utf8]{inputenc}
\usepackage{graphicx}
\usepackage{siunitx}
\usepackage{amsmath}
\usepackage{amsfonts}
\usepackage{amssymb}
\usepackage{graphicx}

\begin{document}

\title{Spontaneous symmetry breaking and Higgs mode: \\
comparing Gross-Pitaevskii and nonlinear Klein-Gordon equations}

\author{M. Faccioli$^{1}$ and L. Salasnich$^{1,2}$}

\address{$^{1}$Dipartimento di Fisica e Astronomia ``Galileo Galilei'', 
Universit\`a di Padova, Via Marzolo 8, 35131 Padova, Italy \\
$^{2}$Istituto Nazionale di Ottica (INO) del Consiglio Nazionale delle 
Ricerche (CNR), Via Nello Carrara 1, 50019~Sesto Fiorentino, Italy}

\date{\today}

\begin{abstract} 
We discuss the mechanism of spontaneous symmetry breaking and 
the elementary excitations for a weakly-interacting Bose 
gas at finite temperature. We consider both the non-relativistic case, 
described by the Gross-Pitaevskii equation, and the relativistic one, 
described by the cubic nonlinear Klein-Gordon equation. 
We analyze similarities and
differences in the two equations and, in particular, in the phase and
amplitude modes (i.e. Goldstone and Higgs modes) of the bosonic 
matter field. We show that the coupling between 
phase and amplitude modes gives rise to a single gapless Bogoliubov spectrum 
in the non-relativistic case. Instead, in the relativistic case the spectrum 
has two branches: one is gapless and the other is gapped. 
In the non-relativistic limit we find that the relativistic spectrum 
reduces to the Bogoliubov one. Finally, as an application of the 
above analysis, we consider the Bose-Hubbard 
model close to the superfluid-Mott quantum phase transition 
and we investigate the elementary excitations of its effective action, 
which contains both non-relativistic and relativistic terms.
\end{abstract}

\maketitle

\section{Introduction} 

The mechanism of spontaneous symmetry breaking is widely used 
to study phase transitions \cite{Huang}. Usually 
the approach introduced by Landau \cite{Landau1,Landau2} for 
second-order phase transitions is adopted, where an order parameter 
is identified and its acquiring a non-zero value corresponds 
to a transition from a disordered phase to an ordered one. 
In other words, when a nonlinearity is added to the symmetric 
problem, and its strength exceeds a critical value, 
there is a loss of symmetry in the system, that is 
called spontaneous symmetry breaking, alias self-trapping into 
an asymmetric state \cite{Malomed}. 
In particular for weakly-interacting Bose gases the 
spontaneous breaking of the $U(1)$ group leads to the transition 
to a superfluid phase \cite{Huang}. In this normal-to-superfluid 
phase transition the order parameter is the mean value of 
the bosonic matter field both in the non-relativistic 
case \cite{Stoof,Atland} and the relativistic one \cite{Kapusta}. 
In the last years, symmetry breaking with the subsequent self-trapping 
has been investigated intensively by our group in the case of 
non-relativistic bosonic and fermionic superfluids made of alkali-metal 
atoms under the action of an external double-well 
potential \cite{Sala1,Sala2} or in the presence of 
Josphson junctions \cite{Sala3,Sala4,Sala5}. 

In this review paper we compute and study the spectrum of 
elementary excitations for both non-relativistic 
and relativistic Bose gases in the ordered phase 
of the normal-to-superfluid phase transition. 
We calculate the elementary excitations by expanding the bosonic 
matter field as the sum of its mean value and fluctuations around it. 
We first study the Euclidean action 
of the bosonic gases and derive the elementary excitations as an intermediate 
step in the computation of the grand canonical potential. 
Then we consider the equations of motion of the bosonic field, 
which are the Gross-Pitaevskii equation \cite{Gross,Pitaevskii} 
in the non-relativistic case 
and the cubic nonlinear Klein-Gordon equation \cite{Klein,Gordon} 
in the relativistic case, 
and we derive the linear equations of motion for fluctuations. We show 
that the complex fluctuating field around the symmetry-breaking 
uniform and constant solution can be written in terms of 
the angle field of the phase, the so-called Goldstone field \cite{Goldstone}
and an amplitude field, the so-called Higgs field \cite{Higgs}. 
For a discussion of Goldstone and Higgs fields in 
Condensed Matter Physics see the recent review Ref. \cite{Varma}. 
We then compare the results found for non-relativistic and 
relativistic cases. {In the last years, 
the interplay between the spectrum of spontaneously 
broken ground state of the relativistic and the non-relativistic 
theories has thoroughly been studied (see, for instance, 
\cite{Leutwyler,Brauner,Endlich,Watanabe}).} 
Here we show that, while the non-relativistic Bose gas 
is characterized only by a gapless (Goldstone-like) mode, 
the relativistic Bose gas has also also a gapped (Higgs-like) mode,  
whose energy gap goes to infinity as the non-relativistic limit 
is approached. {In fact, the appearance of the gapless
spectrum is the direct consequence of the general Goldstone theorem, which 
says that the number of gapless modes is equal to number of the number of 
broken generators \cite{Goldstone}. More precisely, 
as shown by Nielsen and Chadha \cite{Nielsen}, in general there are 
two types of Goldstone bosons: those with an energy proportional 
to an even power of the momentum and those with a dispersion 
relation that is an odd power of the momentum. Within this context, 
a generalized Goldstone theorem holds \cite{Nielsen}: the sum of twice 
the number of Goldstone modes of the first type and the number 
of Goldstone modes of the second type is at least equal to the number of 
independent broken symmetry generators. In our case we find one gapless mode
since the broken symmetry group is $U(1)$, which has only one generator.}

In the last section, the methods used for the weakly-interacting Bose gas 
are used to investigate the Bose-Hubbard model \cite{Sachdev}, which 
describes the non-relativistic dynamics of bosons on a lattice. 
Quite remarkably, close to the superfluid-Mott quantum phase transition, 
the Bose-Hubbard model is captured by an effective action which 
contains both non-relativistic and relativistic terms. We calculate and 
analyze the spectrum of elementary excitations of this effective action. 

\section{Spontaneous symmetry breaking: 
non-relativistic case} 

\subsection{Elementary excitations from non-relativistic 
partition function}

Let us consider a non-relativistic gas of weakly-interacting bosons  
in a volume $V$ at absolute temperature $T$.  
The Eucidean action (imaginary time formalism) of the system is given 
by \cite{Stoof,Atland}:
\begin{equation}
\label{1}
S = \int_{0}^{\hbar \beta} d \tau \ \int_{V}  
d^{D}\vec{r} \; \left \{ \psi^{*}\left(\hbar\frac{\partial}
{\partial \tau} - \frac{\hbar^{2}\nabla^{2}}{2m} - \mu \right)\psi 
+ \frac{g}{2} \vert \psi \vert ^{4}  \right \} 
\end{equation}
where $\psi(\vec{r},\tau)$ is the bosonic matter field, 
$m$ is the mass of each bosonic particle and 
$\mu$ is the chemical potential which fixes the thermal average 
number of bosons in the system. 
We assume that the gas is dilute, such that we can approximate 
the interaction potential $V(\vec{r})$ with a contact interaction, 
i.e. setting $V(\vec{r}) = g \delta(\vec{r})$, where 
the coupling $g$ by construction reads: 
\begin{equation}
g = \int_{V} \, d^{D}\vec{r} \, V(\vec{r})
\end{equation}
The constant $\beta$ is related to the absolute temperature $T$ by: 
\begin{equation}
\beta = \frac{1}{k_{B} T}
\end{equation}
where $k_{B}$ is the Boltzmann constant.
By using functional integration can now define the partition 
function $Z$, and the grand canonical potential $\Omega$, as 
follows \cite{Stoof,Atland}:
\begin{eqnarray}
Z= \int \, D[\psi,\psi^{*}] \, exp\left 
\{ - \frac{S[\psi,\psi^{*}]}{\hbar} \right \} \\
\Omega= -\frac{1}{\beta} ln(Z^{}) \quad . 
\end{eqnarray}
In the Lagrangian we can now consider the effective potential, 
let us call it $V_{eff}$ defined as:
\begin{equation}
V_{eff} = -\mu \vert \psi \vert^{2} + \frac{g}{2} \vert \psi \vert^{4}
\end{equation}
The phase transition correspond to a spontaneous symmetry 
breaking process and for this reason we need to find the minima 
of this potential \cite{Stoof,Atland}. We impose the conditions 
of stationarity on the first derivative:
\begin{eqnarray}
\frac{\partial V_{eff}}{\partial\psi^{*}}= \psi(g\vert 
\psi\vert^{2} - \mu) = 0
\end{eqnarray}
and the minimum is given by:
\begin{equation}
\label{2}
\vert \psi_{0} \vert = \begin{cases}
 0 & \mbox{if}  \;\; \mu  <0 \\ 
\sqrt{\frac{\mu }{g}} & \mbox{if} \;\;  \mu > 0 
\end{cases}
\end{equation}
The superfluid regime corresponds to the lower case. 
It is a condition on the modulus of $\psi$ and therefore we have 
a circle of minima of radius $\vert \psi_{0} \vert$. 
The choice of a particular minimum breaks the $U(1)$ symmetry 
of the Lagrangian. For the superfluid phase we will take the 
real-valued vacuum expectation value, i.e. 
$\psi_{0}=\psi_{0}^{*}=\vert \psi_{0} \vert$. 

To maintain full generality in the following calculations however 
we leave the value of $\psi_{0}$ implicit. 
We can expand $\psi$ as follows:
\begin{equation}
\psi(\vec{r}, \tau) = \psi_{0} + \eta(\vec{r}, \tau)
\end{equation} 
where $\eta$ is the complex fluctuation field. We can now expand 
the Lagrangian to the second-order (i.e. Gaussian) in the fluctuations: 
\begin{widetext}
\begin{multline}
S = \int_{0}^{\hbar\beta} d\tau \, \int_{V} d^{D}\vec{r} \quad  
\big\lbrace -\mu\psi_{0}^{2} +\frac{1}{2}g\psi_{0}^{4} + 
\psi_{0}\hbar\frac{\partial}{\partial\tau}\psi -\mu\psi_{0}
(\eta + \eta^{*}) + g\psi_{0}^{3}(\eta + \eta^{*}) + \\
\eta^{*}\left(\hbar\frac{\partial}{\partial\tau} -
\frac{\hbar^{2}\nabla^{2}}{2m} -\mu + 2g\psi_{0}\right)\eta 
+\frac{g}{2}\psi_{0}^{2}(\eta\eta+\eta^{*}\eta^{*}) \rbrace
\end{multline}
\end{widetext}
The linear terms are written for the sake of completeness 
but they do not contribute. Indeed the linear terms in the 
fluctuations cancel out. Instead the linear terms in the 
derivatives give no contribution to the equation of motion.  

The next step is to expand the fluctuation field in the Fourier space as: 
\begin{equation}
\eta = \sqrt{\frac{1}{V\hbar \beta}} \sum_{n,\vec{q}} 
\eta_{n,\vec{q}} e^{i(\omega_{n}\tau +  \vec{q}\vec{r})} 
\end{equation}
where $\omega_{n}$ are the Matsubara frequencies:
\begin{equation}
\omega_{n} = \frac{2\pi n}{\hbar \beta} \quad n \in \mathbb{Z}
\end{equation}
Let $S_{0}$ be the part of the action which does not depend 
on $\eta$ and $\eta^{*}$. The grand canonical potential 
for the constant term results: 
\begin{equation}
\Omega_{0} = - V\frac{\mu^{2}}{2g} 
\end{equation}
For the quadratic part instead, $S_{2}$, we will use the Fourier 
transform defined above and the fact that:
\begin{eqnarray}
\int_{V} \, d^{D}\vec{r} \quad \frac{1}{V} e^{i(\vec{q} - 
\vec{q'})\vec{r}} = \delta_{\vec{q},\vec{q'}} \\
\int_{0}^{\hbar\beta} \, d\tau \quad \frac{1}{\hbar\beta} 
e^{i(\omega_{n}-\omega_{n'})\tau} = \delta_{n,n'} 
\end{eqnarray}
where $\delta_{\vec{q}\vec{q'}}$ is the Kroenecker delta. 
We can write $S_{2}$ as:
\begin{widetext}
\begin{multline}
S_{2} = \frac{1}{2}\sum_{n,\vec{q}}\sum_{n',\vec{q'}}
\frac{1}{V\hbar \beta}\int_{0}^{\hbar\beta} \ d\tau\int_{V} 
\ d^{D}\vec{r} \, \lbrace e^{i(\omega_{n}-\omega_{n'})\tau 
+ i(\vec{q} - \vec{q'})\vec{r}}\eta_{n',\vec{q'}}^{*}[i\hbar\omega_{n} 
+ \frac{\hbar^{2}q^{2}}{2m} - \mu +2g\psi_{0}^{2}]\eta_{n,\vec{q}}  + \\
 e^{-i(\omega_{n}-\omega_{n'})\tau - i(\vec{q} - 
\vec{q'})\vec{r}}\eta_{-n',-\vec{q}'}^{*}[ -i\hbar\omega_{n} 
+ \frac{\hbar^{2}q^{2}}{2m} - \mu +2g\psi_{0}^{2} ]
\eta_{-n,-\vec{q}} + \\ 
\frac{g}{2}\psi_{0}^{2}(e^{i(\omega_{n}+\omega_{n'})\tau 
+ i(\vec{q} + \vec{q'})\vec{r}}\eta_{n,\vec{q}}\eta_{n',\vec{q'}} 
+ e^{-i(\omega_{n} + \omega_{n'})\tau - i(\vec{q} 
+ \vec{q'})\vec{r}}\eta_{-n,-\vec{q}}\eta_{-n',-\vec{q'}} 
+ \\ e^{-i(\omega_{n}+\omega_{n'})\tau - i(\vec{q} + 
\vec{q'})\vec{r}}\eta_{n,\vec{q}}^{*}\eta_{n',\vec{q'}}^{*} 
+ e^{i(\omega_{n} + \omega_{n'})\tau + i(\vec{q} + 
\vec{q'})\vec{r}}\eta_{-n,-\vec{q}}^{*}\eta_{-n',-\vec{q'}}^{*})\rbrace
\end{multline}
\end{widetext}
Hence using the relations involving the Kroenecker deltas 
written above we can write the precedent equation in a 
different (and far more simple) form, namely involving a 
matrix formalism:
\begin{equation}
S_{2}^{} = \frac{1}{2} \sum_{n,\vec{q}} \begin{bmatrix}
\eta_{n,\vec{q}}^{*} & \eta_{-n,-\vec{q}} 
\end{bmatrix} M^{} \begin{bmatrix}
\eta_{n,\vec{q}} \\
\eta_{-n,-\vec{q}}^{*}
\end{bmatrix}
\end{equation}
where $M$ is the matrix given by:
\begin{equation}
M = \begin{bmatrix}
i\hbar\omega_{n} + \frac{\hbar^{2}q^{2}}{2m} - \mu 
+2g\psi_{0}^{2} & g\psi_{0}^{2}  \\
g\psi_{0}^{2}  & - i\hbar\omega_{n} + \frac{\hbar^{2}
q^{2}}{2m}  - \mu +2g\psi_{0}^{2} 
\end{bmatrix}
\end{equation}
The second-order correction contribution to the partition 
function is given by:
\begin{equation}
Z_{2} = \int \, D[\eta, \eta^{*}] \quad e^{- S_{2}}
\end{equation}
and the second-order correction to the grand canonical potential 
is given by: 
\begin{equation}
\Omega_{2} = - \frac{1}{\beta} ln Z_{2}^{} = \frac{1}{2\beta} 
\sum_{n,\vec{q}} ln(detM^{}) = \frac{1}{2\beta}\sum_{n,\vec{q}} 
ln [\hbar^{2}\omega_{n}^{2} + (E_{q}^{})^{2}]
\end{equation}
where $E_{q}$ reads: 
\begin{equation}
E_{q} = \sqrt{\frac{\hbar^{2}q^{2}}{2m}\left(\frac{\hbar^{2}
q^{2}}{2m} 
- 2\mu + 4g\psi_{0}^{2}\right) + \mu^{2} +3g^{2}\psi_{0}^{4} 
-4g\psi_{0}^{2}\mu}
\end{equation}
After the summation over Matsubara frequencies we can finally write:
\begin{equation}
\Omega_{2}^{} = \sum_{\vec{q}} \left \{ \frac{E_{q}^{}}{2} + 
\frac{1}{\beta}ln(1 - e^{-\beta E_{q}}) \right\} 
\end{equation}
Putting all parts of the grand canonical potential together:
\begin{equation}
\Omega = \Omega_{0} + \Omega_{2}^{(0)} + \Omega_{2}^{(T)}
\end{equation}
where $\Omega_{0}$ is the mean-field potential while 
$\Omega_{2}^{(0)}$, the zero-point energy, and $\Omega_{2}^{(T)}$, 
the thermodynamic fluctuation term, are given by:
\begin{eqnarray}
\Omega_{2}^{(0)} = \sum_{\vec{q}} \frac{E_{q}^{}}{2} \\
\Omega_{2}^{(T)} = \sum_{\vec{q}} \frac{1}{\beta}ln(1 - e^{-\beta E_{q}}) \; .
\end{eqnarray}
For further details on the derivation of these equations and the 
renormalization of the divergent Gaussian grand potential $\Omega_2^{(0)}$ 
see Ref. \cite{Salasnich}. 

It is clear that these calculations hold for both superfluid and normal 
phases: in all calculations we left implicit the value of $\psi_{0}$. 
For the superfluid phase $\psi_{0}^{2} = \frac{\mu}{g}$ and therefore:
\begin{equation}
E_{q} = \sqrt{\frac{\hbar^{2}q^{2}}{2m}\left(\frac{\hbar^{2}
q^{2}}{2m} + 2\mu\right)}
\label{bogo}
\end{equation} 
This spectrum is known as the Bogoliubov spectrum \cite{Bogoliubov} 
of elementary excitations of the non-relativistic Bose gas. 
This is a gapless spectrum and, at small momenta, it becomes 
the phonon mode $E_q \simeq \sqrt{\mu/m} \hbar q$, which can be identified 
as the Goldstone mode that appears necessarily in models exhibiting 
spontaneous breakdown of continuous symmetries \cite{Goldstone}. 
{Thus, the Goldstone mode is only an approximation 
of the Bogoliubov mode. One finds a pure Goldstone mode 
only freezing amplitude fluctuations.}

\subsection{Elementary excitations from non-relativistic 
equation of motion} 

\subsubsection{Non-relativistic complex fluctuations}

By imposing the stationarity condition on the non-relativistic 
action \eqref{1}, 
after having performed a Wick rotation from imaginary time 
to real time, one gets the Gross-Pitaevskii equation 
\cite{Gross,Pitaevskii} 
for a weakly interacting bosonic gas which is given by:
\begin{equation}
\label{3}
i\hbar\frac{\partial}{\partial t}\psi = 
-\frac{\hbar^{2}\nabla^{2}}{2m}\psi -\mu\psi + g\vert \psi \vert^{2}\psi  
\end{equation} 
Let $\psi_{0}$ be the value of the $\psi(\vec{r},\tau)$ field, 
which satisfies the condition \eqref{2}. Let $\eta(\vec{r},\tau)$ 
be the fluctuation around that value. If we expand the 
Gross-Pitaevskii equation to the first order in the fluctuations we obtain: 
\begin{equation}
i\hbar\frac{\partial}{\partial t}\eta = -\frac{\hbar^{2}
\nabla^{2}}{2m}\eta -\mu\eta + g\psi_{0}^{2}(2\eta + \eta^{*})
\end{equation}
If we now perform a Fourier Transform we obtain:
\begin{widetext}
\begin{eqnarray}
(\hbar\omega + \frac{\hbar^{2}q^{2}}{2m} - \mu + 
2g\psi_{0}^{2})\eta_{\omega, \vec{q}}e^{-i\omega t} + 
g\psi_{0}^{2}\eta_{\omega, -\vec{q}}^{*}e^{+i\omega t} = 0 \\
(-\hbar\omega + \frac{\hbar^{2}q^{2}}{2m} - \mu + 
2g\psi_{0}^{2})\eta_{\omega, -\vec{q}}^{*}e^{+i\omega t} + 
g\psi_{0}^{2}\eta_{\omega, \vec{q}}e^{-i\omega t} = 0 
\end{eqnarray}
\end{widetext}
which in turns gives:
\begin{equation}
\hbar^{2}\omega^{2} = \frac{\hbar^{4}q^{4}}{4m^{2}} + 
2\frac{\hbar^{2}q^{2}}{2m}(2g\psi_{0}^{2} -\mu) + \mu^{2} - 
4g\psi_{0}^{2}\mu +3g^{2}\psi_{0}^{4}
\end{equation}
In the superfluid regime, where $\psi_{0}^{2} = \frac{\mu}{g}$, this equation 
gives exactly Eq. (\ref{bogo}). Thus, the spectrum obtained from the 
equation of motion is the same one derived from the partition function.

\subsubsection{Non-relativistic amplitude and 
phase fluctuations} \label{ampl_phase}

We will now compute the spectrum in a slightly different 
way. We now consider separately the phase and the amplitude 
fluctuations. We thus write the boson field $\psi(\vec{r},t)$ 
this time as:
\begin{equation}
\psi(\vec{r},t) = (\psi_{0} + \sigma(\vec{r},t))exp(i\theta(\vec{r},t))
\end{equation}
i.e. including both the amplitude fluctuation field, $\sigma$, 
and the phase fluctuation field, $\theta$, 
the Gross-Pitaevskii equation \eqref{3} 
becomes (using the value of $\psi_{0}$ for the 
superfluid phase) at the first order in $\theta$ and $\sigma$: 
\begin{equation}
i\hbar(\frac{\partial}{\partial t}\sigma + \psi_{0}i\theta) 
= -\frac{\hbar^{2}}{2m}\nabla^{2}(\sigma + i\theta) + 2\mu\sigma
\end{equation}
This equation can be split in its real and imaginary parts. 
The resulting equations are coupled for $\theta$ and $\sigma$:
\begin{eqnarray}
-\hbar\frac{\partial}{\partial t}\psi_{0}\theta + 
\frac{\hbar}{2m}\nabla^{2}\sigma - 2\mu\sigma = 0 \\
\hbar\frac{\partial}{\partial t}\sigma + \frac{\hbar^{2}
\psi_{0}}{2m}\nabla^{2}\theta = 0
\end{eqnarray}
By performing now a Fourier transform we obtain:
\begin{eqnarray}
i\hbar\omega\theta_{\omega,\vec{q}} + \left(\frac{\hbar^{2}q^{2}}{2m} 
+ 2\mu\right)\sigma_{\omega,\vec{q}} = 0 \\
i\hbar\omega\sigma_{\omega,\vec{q}} + \frac{\hbar^{2}q^{2}}{2m}
\theta_{\omega,\vec{q}} = 0
\end{eqnarray}
where $\theta_{\omega,\vec{q}}$ and $\sigma_{\omega,\vec{q}}$ 
are the Fourier transforms of the fluctuation fields. 
If we substitute, for example, the expression of $\sigma_{\omega,\vec{q}}$ 
obtained by second equation in the first we get:
\begin{equation}
\left[\hbar^{2}\omega^{2} - \left(\frac{\hbar^{2}q^{2}}
{2m} + 2\mu\right)\frac{\hbar^{2}q^{2}}{2m}\right]
\theta_{\omega,\vec{q}} = 0
\end{equation}
solving for $\omega$ we find again the Bogoliubov spectrum, 
Eq. (\ref{bogo}), that is the same results obtained with the other two methods.
Note that if we consider only the phase fluctuations, 
i.e., we impose $\sigma=0$, the Gross-Pitaevskii equation 
in the first order in $\theta$ becomes:
\begin{equation}
i\hbar\frac{\partial}{\partial t}\theta = \frac{\hbar}{2m}\nabla^{2}\theta
\end{equation}
and if we consider the Fourier transform we obtain the following spectrum:
\begin{equation}
\hbar\omega = \frac{\hbar q^{2}}{2m}
\end{equation}
this is a gapless spectrum which has the form of a 
free particle spectrum. Conversely if we consider the 
case $\theta=0$, i.e. we consider only the amplitude 
fluctuations, we get the equation:
\begin{equation}
i\hbar\frac{\partial}{\partial t}\sigma = -\frac{\hbar^{2}}
{2m}\nabla^{2}\sigma + 2\mu\sigma
\end{equation}
which gives the spectrum:
\begin{equation}
\hbar\omega = 2\mu + \frac{\hbar^{2}q^{2}}{2m}
\end{equation}
this time we have a gapped spectrum, the gap being $2\mu$. 
This is consistent with what we would expect by the spontaneous 
symmetry mechanism: the breaking of the $U(1)$ symmetry in fact 
produces always a gapless mode, which is usually called Goldstone mode 
\cite{Goldstone}, and a gapped mode, which in Condensed Matter Physics 
is referred as Higgs mode \cite{Higgs,Varma}. 

\section{Spontaneous Symmetry Breaking: relativistic case} 
\label{Rel}

\subsection{Elementary excitations from relativistic partition function}

Working with the same approximation for the dilute gas 
as in the previous section, for a weakly-interacting 
relativistic gas the Eucidean action is given 
by \cite{Kapusta,Kapusta2,Bernstein,Alford2}: 
\begin{widetext}
\begin{multline}
\label{4}
S = \int_{0}^{\hbar\beta} \, d\tau \int_{V} \,d^{D}\vec{r} 
\quad \Big( \frac{\hbar^{2}}{mc^{2}}\vert\frac{\partial}
{\partial\tau}\psi\vert^{2} + 2\hbar\frac{\mu_{r}}{mc^{2}}
\psi^{*}\frac{\partial}{\partial\tau}\psi + \frac{\hbar^{2}}
{m}\vert \nabla\psi \vert^{2} +   (\frac{\mu_{r}^{2}}{mc^{2}} 
- mc^{2}) \vert \psi \vert^{2} + \frac{g}{2} \vert \psi \vert^{4} \Big) 
\end{multline}
\end{widetext}
where $\psi(\vec{r},\tau)$ is the bosonic matter field 
and we have introduced the relativistic chemical potential, 
$\mu_{r}$ which is given by:
\begin{equation}
\label{8}
\mu_{r} = \mu + mc^{2}
\end{equation}
If we define again an effective potential $V_{eff}$ such as:
\begin{equation}
V_{eff} = -(\frac{\mu_{r}^{2}}{mc^{2}}-mc^{2})\vert \psi 
\vert^{2} + \frac{g}{2} \vert \psi \vert^{4} 
\end{equation}
Clearly if $\mu_{r}^{2}-m^{2}c^{4} > 0$ we have the superfluid 
phase: the $U(1)$ symmetry is broken and therefore we can proceed 
as we have done in the previous section. In particular the minima 
are given by: 
\begin{equation}
\label{5}
\vert \psi_{0} \vert = \begin{cases}
0 & if \quad \mu_{r}^{2}-m^{2}c^{4} < 0 \\
\sqrt{\frac{\frac{\mu_{r}^{2}}{mc^{2}}-mc^{2}}{g}} & if 
\quad \mu_{r}^{2}-m^{2}c^{4} > 0
\end{cases}
\end{equation}
The first case correspond to the normal phase, characterized 
by a mean value of order parameter equal to zero. For both 
phases we choose the real-valued vacuum, let us call it 
$\psi_{0}$. Let us now call $\eta(\vec{r},\tau)$ the fluctuations 
around the minimum. We expand now the action to the second order 
in the fluctuations, maintaining for generality the value of 
$\psi_{0}$ implicit. We obtain:
\begin{widetext}
\begin{multline}
S = V\hbar \beta (-(\frac{\mu_{r}^{2}}{mc^{2}}-mc^{2})
\psi_{0}^{2} + \frac{g}{2}\psi_{0}^{4}) + \\ 
\int_{0}^{\hbar\beta} \, d\tau \int_{V} \,d^{D}\vec{r} 
\quad \lbrace \hbar\frac{\mu_{r}}{mc^{2}} (\eta^{*}
\frac{\partial}{\partial\tau}\eta - \eta\frac{\partial}
{\partial\tau}\eta^{*}) + \frac{\hbar^{2}}{mc^{2}}
\vert\frac{\partial}{\partial\tau}\eta\vert^{2} + 
\frac{\hbar^{2}}{m}\vert \nabla\eta \vert^{2} - \\ 
(\frac{\mu_{r}^{2}}{mc^{2}}-mc^{2})\vert \eta \vert^{2} +
\frac{g}{2}\psi_{0}^{2}(\eta\eta + \eta^{*}\eta^{*} + 
4\vert \eta \vert^{2}) \rbrace
\end{multline}
\end{widetext}
The constant term:
\begin{equation}
S_{0} =  V\hbar \beta(-(\frac{\mu_{r}^{2}}{mc^{2}}-mc^{2})
\psi_{0}^{2} + \frac{g}{2}\psi_{0}^{4}) 
\end{equation}
gives a contribution to the grand canonical potential:
\begin{equation}
\Omega_{0}^{} = V(-(\frac{\mu_{r}^{2}}{mc^{2}}-mc^{2})
\psi_{0}^{2} + \frac{g}{2}\psi_{0}^{4}) 
\end{equation}
whereas the second-order correction of the action can be 
written in a matrix form in the Fourier space (the sum over 
the index $n$ refers to the sum over the Matsubara frequencies): 
\begin{equation}
S_{2} = \frac{1}{2}\sum_{n, \vec{q}} \begin{bmatrix}
\eta_{n,\vec{q}}^{*} & \eta_{-n,-\vec{q}} 
\end{bmatrix} \frac{1}{mc^{2}}M \begin{bmatrix}
\eta_{\vec{q}} \\
\eta_{-\vec{q}}^{*}
\end{bmatrix}
\end{equation}
where M is given by: 
\begin{equation}
M = \begin{bmatrix}
A & B \\
B & C
\end{bmatrix}
\end{equation}
where:
\begin{widetext}
\begin{eqnarray}
A = \hbar^{2}\omega_{n}^{2} + 2\hbar\omega_{n}\mu_{r} +
\hbar^{2}c^{2}q^{2} -(\mu_{r}^{2} - m^{2}c^{4}) + 
2g\psi_{0}^{2}mc^{2}\ \\
B = g\psi_{0}^{2}mc^{2}\\
C =  \hbar^{2}\omega_{n}^{2} - 2\hbar\omega_{n}\mu_{r} +
\hbar^{2}c^{2}q^{2} - (\mu_{r}^{2} - m^{2}c^{4}) + 
2g\psi_{0}^{2}mc^{2}
\end{eqnarray}
\end{widetext}
The second order contribution to the grand canonical 
potential then results: 
\begin{eqnarray}
\Omega_{2} &=& \frac{1}{2\beta} \sum_{n,\vec{q}} 
ln(\frac{1}{m^{2}c^{4}}detM) 
\nonumber 
\\
&=& \frac{1}{2\beta}
\sum_{n,\vec{q}}\sum_{j=\pm} ln [\frac{1}{m^{2}c^{4}}
(\hbar^{2}\omega_{n}^{2} + E_{j,q}^{2})]
\end{eqnarray}
where $E_{\pm,q}$ is given by:
\begin{widetext}
\begin{multline}
E_{\pm,q}^{2} = \hbar^{2}c^{2}q^{2} + 
(\mu_{r}^{2}+ m^{2}c^{4} + 2g\psi_{0}^{2}mc^{2}) 
\pm \sqrt{4\mu_{r}^{2}(\hbar^{2}c^{2}q^{2} + 
m^{2}c^{4} + 2g\psi_{0}^{2}mc^{2}) + g^{2}\psi_{0}^{4}m^{2}c^{4}}
\end{multline}
\end{widetext}
Summing over the Matsubara frequencies we can finally write:
\begin{equation}
\Omega_{2} = \sum_{\vec{q}}\sum_{j=\pm} \left 
\{\frac{E_{q,j}}{2} + \frac{1}{\beta}ln(1 - 
e^{-\beta E_{q,j}}) \right \}
\end{equation}
Putting all terms of the grand canonical potential 
density together we obtain:
\begin{equation}
\Omega = \Omega_{0} + \Omega_{2}^{(0)} + \Omega_{2}^{(T)}
\end{equation}
where $\Omega_{2}^{(0)}$, the zero-point Gaussian grand 
canonical potential density, and $\Omega_{2}^{(T)}$ is 
the fluctuation term, defined respectively as:
\begin{eqnarray}
\Omega_{2}^{(0)} = \sum_{\vec{q}}\sum_{j=\pm} \frac{E_{q,j}}{2} \\
\Omega_{2}^{(T)} =  \sum_{\vec{q}}\sum_{j=\pm} \frac{1}
{\beta}ln(1 - e^{-\beta E_{q,j}}) 
\end{eqnarray}
For the superfluid phase $\psi_{0}^{2}=\frac{\frac{\mu_{r}^{2}}{mc^{2}} 
- mc^{2}}{g}$ and therefore the spectrum becomes:
\begin{equation}
\label{7}
E_{\pm,q}^{2} = \hbar^{2}c^{2}q^{2} + (3\mu_{r}^{2} - 
m^{2}c^{4}) \pm \sqrt{4\mu_{r}^{2}\hbar^{2}c^{2}q^{2} + 
(3\mu_{r}^{2} - m^{2}c^{4})^{2}}
\end{equation}

\subsection{Elementary excitations from nonlinear Klein-Gordon 
equation}

\subsubsection{Relativistic complex fluctuations}

By extremizing the relativistic action action \eqref{4}, after having performed 
the Wick rotation as in the non-relativistic case, 
we find the cubic nonlinear Klein-Gordon equation \cite{Klein,Gordon} 
for a bosonic gas with relativistic chemical potential $\mu_{r}$
\cite{Kapusta,Kapusta2,Bernstein,Alford2}:
\begin{equation}
\label{6}
(\frac{\hbar^{2}}{m}D_{\nu}D^{\nu} + m c^{2} + 
g \vert \psi \vert^{2})\psi = 0
\end{equation}
where $D_{\nu}$ is the covariant derivative defined by:
\begin{eqnarray}
D_{0} = \frac{1}{c}\frac{\partial}{\partial t} - i\frac{\mu_{r}}{\hbar c} \\
D_{i} = \partial_{i}
\end{eqnarray}
so the nonlinear Klein-Gordon equation can be written:
\begin{equation}
(\hbar^{2}\partial_{t}^{2} - 2i\hbar\mu_{r}\partial_{t} - 
\hbar^{2}c^{2}\nabla^{2} - (\mu_{r}^{2} - m^{2}c^{4}) + 
g mc^{2}\vert \psi \vert^{2})\psi = 0
\end{equation}
We now write the field as the sum of the vacuum expectation 
value, $\psi_{0}$, and a fluctuation field, let us call 
it $\eta$. The nonlinear Klein-Gordon equation in the first-order 
of the fluctuation is given by:
\begin{widetext}
\begin{equation}
(\hbar^{2}\partial_{t}^{2} - 2i\hbar\mu_{r}\partial_{t} - 
\hbar^{2}c^{2}\nabla^{2} - (\mu_{r}^{2} - m^{2}c^{4}) + 
2g mc^{2}\psi_{0}^{2})\eta + g mc^{2}\psi_{0}^{2}\eta^{*} = 0
\end{equation}
\end{widetext}
We now perform a Fourier transform for this equation and 
its complex conjugate, obtaining:
\begin{widetext}
\begin{eqnarray}
(-\hbar^{2}\omega^{2} - 2\mu_{r}\hbar \omega + 
\hbar^{2}c^{2}q^{2} - (\mu_{r}^{2} - m^{2}c^{4}) 
+ 2g mc^{2}\psi_{0}^{2})\eta_{\omega, \vec{q}}e^{-i\omega t} 
+ g mc^{2}\psi_{0}^{2}\eta_{\omega, -\vec{q}}^{*}e^{+i\omega t} = 0 \\
(-\hbar^{2}\omega^{2} + 2\mu_{r} \hbar\omega + \hbar^{2}c^{2}q^{2} 
- (\mu_{r}^{2} - m^{2}c^{4}) + 2g mc^{2}\psi_{0}^{2})
\eta_{\omega, -\vec{q}}^{*}e^{+i\omega t} + g mc^{2}
\eta_{\omega, \vec{q}}e^{-i\omega t} = 0
\end{eqnarray}
\end{widetext}
These equation give the following solutions:
\begin{widetext}
\begin{multline}
\hbar^{2}\omega_{\pm}^{2} = \hbar^{2}c^{2}q^{2} + m^{2}c^{4} 
+ \mu_{r}^{2} + 2g mc^{2}\psi_{0}^{2} \pm \sqrt{4\mu_{r}^{2}
(\hbar^{2}c^{2}q^{2} + m^{2}c^{4}  + 2g mc^{2}\psi_{0}^{2}) 
+ g^{2} m^{2}c^{4}\psi_{0}^{4}}
\end{multline}
\end{widetext}
and substituting the value of $\psi_{0}$ for the superfluid phase
given by Equation \eqref{5} we obtain:
\begin{equation}
\hbar^{2}\omega_{\pm}^{2} = \hbar^{2}c^{2}q^{2} + 
(3\mu_{r}^{2} - m^{2}c^{4}) \pm \sqrt{4\mu_{r}^{2}
\hbar^{2}c^{2}q^{2} + (3\mu_{r}^{2} - m^{2}c^{4})^{2}}
\end{equation}
Also in the relativistic case we have the same spectrum 
found using the partition function.

\subsubsection{Relativistic amplitude and phase fluctuations}

We show now that we can find the spectrum also by expanding 
the matter field $\psi$ as:
\begin{equation}
\psi = (\psi_{0} + \sigma(\vec{r},t))exp(i\theta(\vec{r},t))
\end{equation}
where $\sigma$ is the Higgs amplitude field and 
$\theta(\vec{r},t)$ is the Goldstone angle field. 
Using again the value of $\psi_{0}$ for the superfluid phase given by
the condition \eqref{5},
we obtain by expanding the cubic nonlinear Klein-Gordon Equation \eqref{6}  
in the first order of the fluctuations:
\begin{widetext}
\begin{equation}
(\hbar^{2}\partial_{t}^{2} - 2i\hbar\mu_{r}\partial_{t} -
\hbar^{2}c^{2}\nabla^{2})(\sigma + i\psi_{0}\theta) + 
2(\mu_{r}^{2} - m^{2}c^{4})\sigma = 0
\end{equation}
\end{widetext}
which like the non-relativistic case can be decoupled in 
its imaginary and real parts. The equations are however coupled:
\begin{eqnarray}
(\hbar^{2}\partial_{t}^{2} - \hbar^{2}c^{2}\nabla^{2} 
+ 2(\mu_{r}^{2} - m^{2}c^{4}))\sigma +  2\hbar\mu_{r}
\partial_{t}\psi_{0}\theta= 0 \\
(\hbar^{2}\partial_{t}^{2} -\hbar^{2}c^{2}\nabla^{2})
\psi_{0}\theta -2\hbar\mu_{r}\partial_{t}\sigma = 0
\end{eqnarray}
By performing the Fourier transform we obtain:
\begin{eqnarray}
(-\hbar^{2}\omega^{2} - \hbar^{2}c^{2}q^{2} + 
2(\mu_{r}^{2} - m^{2}c^{4}))\sigma_{\omega,\vec{q}} 
-  2\omega\hbar\mu_{r}\psi_{0}\theta_{\omega,\vec{q}}= 0 \\
(-\hbar^{2}\omega^{2} - \hbar^{2}c^{2}q^{2})\psi_{0}
\theta_{\omega,\vec{q}} + 2\omega\hbar\mu_{r}\sigma_{\omega,\vec{q}} = 0
\end{eqnarray}
where $\theta_{\omega,\vec{q}}$ and $\sigma_{\omega,\vec{q}}$ 
are the Fourier transforms of the fluctuation fields. 
By using the expression of $\sigma_{\omega,\vec{q}}$ 
found by solving the second equation and substituting 
it in the first we obtain:
\begin{widetext}
\begin{multline}
\lbrace[(-\hbar^{2}\omega^{2} - \hbar^{2}c^{2}q^{2} 
+ 2(\mu_{r}^{2} - m^{2}c^{4}))(-\hbar^{2}\omega^{2} 
- \hbar^{2}c^{2}q^{2} + 2(\mu_{r}^{2} - m^{2}c^{4})]  - 
4\omega^{2}\hbar^{2}\mu_{r}^{2}\rbrace\psi_{0}\theta_{\omega,\vec{q}}= 0
\end{multline}
\end{widetext}
which gives:
\begin{widetext}
\begin{multline}
\hbar^{4}w^{4} - 2\hbar^{2}\omega^{2}(\hbar^{2}c^{2}q^{2} 
+ 3\mu_{r}^{2} - m^{2}c^{4}) + \hbar^{2}c^{2}q^{2}
[\hbar^{2}c^{2}q^{2} - 2(\mu_{r}^{2} - m^{2}c^{4})] = 0
\end{multline}
\end{widetext}
and $\omega$ is therefore given by:
\begin{equation}
\hbar\omega_{\pm} = \sqrt{\hbar^{2}c^{2}q^{2} + 
(3\mu_{r}^{2} - m^{2}c^{4}) \pm \sqrt{\hbar^{2}c^{2}q^{2} 
+ (3\mu_{r}^{2} - m^{2}c^{4})^{2}}}
\end{equation}
which is exactly the same result found with the other methods. 

It is important to observe that the cubic nonlinear Klein-Gordon 
equation is used to describe not only a relativistic Bose gas 
but also the dynamics of Cooper pairs in BCS 
superconductors \cite{Varma,Cea}.  

\section{Analysis and comparison of spectra} \label{AC}

We now proceed to study the spectra we have found.
In the non-relativistic case we have found a gapless Bogoliubov 
spectrum, given by:
\begin{equation}
\label{10}
\hbar\omega = \sqrt{\frac{\hbar^{2}q^{2}}{2m}
\left(\frac{\hbar^{2}q^{2}}{2m} + 2\mu\right)}
\end{equation} 
This spectrum for small momenta gives:
\begin{eqnarray}
\hbar\omega \simeq \sqrt{\frac{\mu}{m}}\hbar q &  if & 
\frac{\hbar^{2} q^{2}}{2m} \ll \mu
\end{eqnarray}
Therefore for small momenta we obtain a phonon-like linear
spectrum.
For sufficiently large momenta we instead get: 
\begin{eqnarray}
\hbar\omega_{q} \simeq \frac{\hbar^{2}q^{2}}{2m} & if & 
\frac{\hbar^{2} q^{2}}{2m} \gg \mu
\end{eqnarray} 
In this case we obtained a free-particle quadratic spectrum.
This shows that the contact interaction does not affect the spectrum
in the limit of high energies, whereas in the opposite limit we get
a linear spectrum.
\\
We now consider the relativistic spectrum. In the calculations we found 
two modes, namely:
\begin{widetext}
\begin{equation}
\hbar\omega_{\pm} = \sqrt{\hbar^{2}c^{2}q^{2} + (3\mu_{r}^{2} 
- m^{2}c^{4}) \pm \sqrt{4\mu_{r}^{2}\hbar^{2}c^{2}q^{2} + 
(3\mu_{r}^{2} - m^{2}c^{4})^{2}}}
\end{equation}
\end{widetext}
At high energies we have that the term involving the 
higher-degree momentum becomes dominant and modes are given by:
\begin{equation}
\label{ultra}
\hbar\omega_{\pm} = \hbar c q  
\end{equation}  
In this limit we have two free relativistic particles spectra:
as in the non-relativistic case we obtained that at high energies 
the spectra are unaffected by the contact interaction. We not that 
we have two modes corresponding to a particle and its antiparticle.
For small momenta the situation is different. In fact using the 
relation between the relativistic chemical potential and the 
non-relativistic one \eqref{8},
and the fact that $\mu \ll mc^{2}$ the Taylor expansion 
around $q \rightarrow 0$ yields:
\begin{eqnarray}
\hbar\omega_{-} = \sqrt{\frac{\mu}{m}}\hbar q \\
\hbar\omega_{+} = 2m c^{2} + \frac{\hbar^{2} q^{2}}{2m}
\end{eqnarray}
We note that we have two modes: a gapless, 
i.e. Goldstone for the relativistic case mode, 
which is linear for small momenta like the Bogoliubov spectrum 
in the same limit, and a gapped mode, i.e. the Higgs mode 
for the relativistic case. Therefore, as expected, from the spontaneous 
symmetry breaking 
of the $U(1)$ symmetry we find the presence of both Goldstone 
and Higgs modes. The gapped mode for small energies is given by 
the sum between the gap, and a quadratic term in the momenta 
which has the form of the spectrum of a non-relativistic 
free particle (which corresponds to the high momenta limit 
in the non-relativistic case), whereas the gapless mode in the 
same limit is actually the same.  

Until now, however, we have not recovered the Bogoliubov spectrum. 
Let us now consider again the Goldstone (relativistic) mode, $\hbar\omega_{-}$. 
For small momenta it can be written as:
\begin{widetext}
\begin{multline}
\label{bogo2}
\hbar\omega_{-} =  \sqrt{\hbar^{2}c^{2}q^{2} - 
\frac{2\mu_{r}^{2}\hbar^{2}c^{2}q^{2}}{3\mu_{r}^{2} - 
m^{2}c^{4}} + \frac{4\mu_{r}^{4}\hbar^{4}c^{4}q^{4}}
{(3\mu_{r}^{2} - m^{2}c^{4})^{3}}} = 
\sqrt{\frac{\hbar^{2}c^{2}q^{2}}{3\mu_{r}^{2} - m^{2}c^{4}}
\left(\frac{4\mu_{r}^{2}\hbar^{2}c^{2}q^{2}}{(3\mu_{r}^{2} - m^{2}c^{4})^{2}} 
+ (\mu_{r}^{2} - m^{2}c^{4})\right)}
\end{multline}
\end{widetext}
and now since we are interested in the non-relativistic case
by imposing $\mu \ll mc^{2}$ we obtain the Bogoliubov spectrum \eqref{10}.
It should also be noted that the Bogoliubov mode is not, 
actually, the Goldstone mode of the non-relativistic case, 
as we have seen in Section \ref{ampl_phase}. 
In that case the Goldstone mode coincides with the phase mode. 
Similarly the Higgs mode 
corresponds to the amplitude mode. In the relativistic case, 
however, both these modes are relative 
to the total fluctuation around the value of the minimum. 

\section{Application: the Bose-Hubbard model}

An interesting application of the above considerations is the Bose-Hubbard 
model. The model was first introduced by Gersch and Knollman \cite{Gersch} 
as a bosonic version of the Hubbard model for fermions 
on a lattice \cite{Hubbard}. 
The Bose-Hubbard model is used to describe an interacting Bose gas 
confined in a periodic lattice by an external potential. 
We assume that for each site of the lattice the value 
of the potential is the same. 
With this assumption, the bosonic system is described by the 
Hamiltonian \cite{Sachdev}:
\begin{equation}
\hat{H}_{BH} = -J\sum_{\langle ij \rangle}\hat{a}_{i}^{+}\hat{a}_{j} 
- (\mu - \epsilon) \sum_{i} \hat{a}_{i}^{+}\hat{a}_{i} + \frac{U}{2} 
\sum_{i}\hat{a}_{i}^{+}\hat{a}_{i}^{+}\hat{a}_{i}\hat{a}_{i}
\end{equation}
where $\hat{a}_{i}$ is the annihilation operator for the site $i$, 
$\mu$ is the chemical potential of the gas, $J$ is the coupling of 
the interaction between the nearest-neighbors (also called the "hopping" 
term), $\epsilon$ is the energy of the energy of each particle of every 
site due to its kinetic energy and to the confining potential, and 
finally $U$ is proportional to the interaction strength of bosons. 
The Bose-Hubbard model has a phase transition between 
an insulating phase, called the Mott insulating phase, and a superfluid phase.
In particular for a system of $T=0$ and volume $V$ 
for the regions of phase space near the phase transitions,
it can be shown that the behavior of the system is described,
using a RPA approximation treating the hopping term as a perturbation,
by the following action in the imaginary time 
formalism \cite{Sachdev,Dupuis}:
\begin{widetext}
\begin{equation}
S_{BH}^{(RPA)} =  \int_{\mathbb{R}} d\tau \int_{V} d\vec{r} 
\lbrace K_{1}\psi^{*}\frac{\partial}{\partial \tau}\psi + 
K_{2}\vert \frac{\partial}{\partial \tau}\psi \vert^{2} + 
K_{3}\vert \nabla\psi \vert^{2} + c_{2}\vert \psi \vert^{2} + 
c_{4}\vert \psi \vert^{4} \rbrace
\label{mistic}
\end{equation}
\end{widetext}
where $\psi(\vec{r},\tau)$ is an appropriately chosen order parameter 
(related to the mean value of the annihilation operator). 
The coefficients $K_1$, $K_2$, $K_3$, $c_2$, $c_4$ depend on 
the Bose-Hubbard parameters $J$, $\mu$, $\epsilon$, $U$. 
This dependence in shown and discussed in Ref. \cite{Dupuis}. 
The form of the effective action (\ref{mistic}) is strikingly similar 
to the one found  for the relativistic case \eqref{4} due to the 
term $K_{2}\vert \frac{\partial}{\partial \tau}\psi \vert^{2}$. 
Also the non-relativistic term 
$K_{1}\psi^{*}\frac{\partial}{\partial \tau}\psi$
has a correspondence in that action to the term linear in
the relativistic chemical potential.
The phase transition occurs at 
the change of sign of the coefficient of the quadratic term. 
Note that the transition is purely quantum since we are working 
at zero temperature. 
In fact following the same reasoning used in Section \ref{Rel}, 
we find that the minima of the effective potential are given by:
\begin{equation}
\vert \psi_{0} \vert = \begin{cases}
0 & \text{if} \quad c_{2} > 0 \\
\sqrt{\frac{2c_{2}}{c_{4}}} & \text{if} \quad c_{4} < 0
\end{cases}
\end{equation}  
and as before we choose the real-valued minimum for the superfluid phase.
We now write the order parameter as a sum of its mean value and the 
fluctuations:
\begin{equation}
\psi(\vec{r}, \tau) = \psi_{0} + \eta(\vec{r},\tau) 
\end{equation}
and we expand the action up to the second order in the fluctuations and by 
following the same steps of Section \ref{Rel}, we find the spectrum:
\begin{widetext}
\begin{equation}
E_{\pm} = \sqrt{K_{3}q^{2} + \left(\frac{K_{1}^{2}}{2K_{2}} + c_{2} 
+ 4c_{4}\psi_{0}^{2}\right) \pm  \sqrt{\frac{K_{1}^{2}}{K_{2}}K_{3}q^{2} 
+ \frac{K_{1}^{4}}{4K_{2}^{2}} + \frac{K_{1}^{2}}{K_{2}}(c_{2} 
+ 4c_{4}\psi_{0}^{2}) + 4c_{4}^{2}\psi_{0}^{4} }}
\end{equation}
\end{widetext}
and substituting the value of $\psi_{0}$ for the superfluid phase we obtain:
\begin{equation}
\label{BH}
E_{\pm}= \sqrt{K_{3}q^{2} + \left(\frac{K_{1}^{2}}{2K_{2}} - c_{2}\right) 
\pm \sqrt{\frac{K_{1}^{2}}{K_{2}}K_{3}q^{2} + \left(\frac{K_{1}^{2}}{2K_{2}} 
- c_{2}\right)^{2}}}
\end{equation}
This spectrum has the same form of the one found for the superfluid phase
for the relativistic gas \eqref{7}. To better note the formal analogy the
following identifications should be considered:
\begin{eqnarray*}
\hbar^{2} c^{2} \leftrightarrow K_{3} \\
 \mu_{r}^{2} \leftrightarrow \frac{K_{1}^{2}}{2K_{2}} \\
 \mu_{r}^{2} - m^{2} c^{4} \leftrightarrow c_{2}
\end{eqnarray*}
As mentioned above, this formal analogy is possible thanks to the 
inclusion of the relativistic chemical potential in the relativistic 
action \eqref{4}, that gives rise to the linear term in the time derivative 
and a correction to the quadratic term.
\\
It is interesting now to study the behavior of the two modes of 
the spectrum \eqref{BH}
in the limits of low and high momenta.
In particular in the first limit we find at leading order: 
\begin{eqnarray}
E_{+} = \sqrt{2\left(\frac{K_{1}^{2}}{2K_{2}}- c_{2}\right)} + 
\frac{\frac{K_{1}^{2}}{2K_{2}}-c_{2}}{\sqrt{2}\left(\frac{K_{1}^{2}}
{2K_{2}}-c_{2}\right)^{\frac{3}{2}}}K_{3}q^{2} \\
E_{-} = \frac{\sqrt{-c_{2}}}{\sqrt{\frac{K_{1}^{2}}{2K_{2}}-c_{2}}}K_{3}q^{2}
\end{eqnarray} 
We have like in the relativistic gas a gapped Higgs mode ,which for low 
energies is quadratic in momenta, and a gapless Goldstone mode, 
which in the same limit is linear. If we expand up to the next to leading 
order the gapless mode, we obtain:
\begin{equation}
E_{-} = \sqrt{\frac{K_{3}q^{2}}{2\left( \frac{K_{1}^{2}}{2K_{2}} - c_{2} \right)}
\left[ \frac{\left(\frac{K_{1}^{2}}{K_{2}}\right)^{2}K_{3}q^{2}}{2\left( 
\frac{K_{1}^{2}}{2K_{2}} - c_{2} \right)^{2}} - 2c_{2} \right]}
\end{equation}
which is reminiscent of the Bogoliubov spectrum \eqref{bogo}. 
In particular using the identifications written above, 
we obtain the result found in Equation \eqref{bogo2}.
Finally we note that for high momenta we have:
\begin{equation}
E_{\pm} = \sqrt{K_{3}}q
\end{equation}
which is analogous to the relativistic free particle spectrum 
found in Equation \eqref{ultra} for the relativistic superfluid.

\section{Conclusions}

In this brief review we have derived and studied the spectrum of the 
superfluid phase of both non-relativistic and relativistic 
bosonic gases. This phase is described by a spontaneous 
symmetry breaking process of the $U(1)$ group symmetry of the action. 
We have found, in agreement with the expectations, that in both 
cases there is indeed a gapless Goldstone mode due to phase 
fluctuation and a gapped Higgs mode due to amplitude fluctuations. 
However, while in the non-relativistic case the coupling between 
phase and amplitude gives rise to a total gapless Bogoliubov spectrum, 
in the relativistic case both modes are 
possible oscillations modes of the total fluctuation 
around the solution with broken symmetry. 
{The difference between the Goldstone mode and the 
Bogoliubov mode in the non-relativistic case can be interpreted
by noting that in this regime there is not the particle-antiparticle pair
typical of the relativistic case.} Then, we have verified that 
the Bogoliubov spectrum can be obtained as the non-relativistic 
limit of the relativistic Goldstone mode. 
Finally, we have analyzed the Bose-Hubbard model, 
that is characterized by the effective action close to the critical point 
of the Superfluid-Mott quantum phase transition which contains both 
non-relativistic and relativistic terms \cite{Sachdev}. 
Apart some theoretical \cite{Dupuis} and experimental \cite{Endres}
results for the Bose-Hubbard model, the properties of phase and 
amplitude fluctuations in this exotic effective action 
are not yet fully explored.


\begin{thebibliography}{999}

\bibitem{Huang} Huang, K. {\it Statistical Machanics}; 
Wiley and Sons: Hoboken, USA, 1987. 
    
\bibitem{Landau1} Landau, L. Theory of phase transformations. I.  
\emph{Zh. Eksp. Teor. Fiz.} \textbf{1937}, \emph{7}, 19. 

\bibitem{Landau2} Landau, L. Theory of phase transformations. II.  
\emph{Zh. Eksp. Teor. Fiz.} \textbf{1937}, \emph{7}, 627. 

\bibitem{Malomed} Malomed, B.A. {\it Spontaneous Symmetry Breaking, 
Self-Trapping, and Josephson Oscillations}; Springer: Berlin, Germany, 2003. 
  
\bibitem{Stoof} Stoof, H.T.C.; Gubbels, K.B.; Dickerscheid, D.B.M. 
{\it Ultracold Quantum Fields}; Springer: Berlin, Germany, 2009.
  
\bibitem{Atland} Altland A.; Simons B. {\it Condensed Matter 
Field Theory}; Cambridge University Press: Cambridge, UK, 2010.
  
\bibitem{Kapusta} Kapusta, J.I.; Gale, C. \textit{Finite temperature 
field theory. Principles and applications 2nd edition};  
Cambridge University Press: Cambridge, UK, 2006. 

\bibitem{Sala1} Salasnich, L.; Parola, A.; Reatto, L.  
Bose condensate in a double-well trap: Ground state and elementary 
excitations. \emph{Phys. Rev. A} \textbf{1999}, \emph{60}, 4171.

\bibitem{Sala2} Adhikari, S.K.; BA Malomed, B.A.; Salasnich, L.; Toigo, F. 
Spontaneous symmetry breaking of Bose-Fermi mixtures 
in double-well potentials. 
\emph{Phys. Rev. A} \textbf{2010}, \emph{81}, 053630.

\bibitem{Sala3} Mazzarella, G.; Salasnich, L.; Salerno, M.; Toigo, F.  
Atomic Josephson junction with two bosonic species.  
\emph{J. Phys. A: At.Mol.Opt.Phys.} \textbf{2009}, \emph{42}, 125301.

\bibitem{Sala4} Mazzarella, G.; Salasnich, L. 
Spontaneous symmetry breaking and collapse in bosonic 
Josephson junctions. \emph{Phys. Rev. A} \textbf{2010}, \emph{82}, 033611.

\bibitem{Sala5} Chen, Z.; Li, Y.; Malomed, B.A.; Salasnich, L. 
Spontaneous symmetry breaking of fundamental states, vortices, and dipoles
in two- and one-dimensional linearly coupled traps with cubic
self-attractions. \emph{Phys. Rev. A} \textbf{2010}, \emph{82}, 033611.

\bibitem{Gross} Gross, E.P. Structure of a quantized vortex in 
boson systems. \emph{Nuovo Cimento} \textbf{1961}, \emph{20}, 3.
  
\bibitem{Pitaevskii} Pitaevskii, L.P. 
Vortex lines in an imperfect Bose gas.  
\emph{Sov. Phys. JETP} \textbf{1961}, \emph{1}, 2.

\bibitem{Klein} Klein, O. Quantentheorie und funfdimensionale 
Relativitatstheorie, \emph{Z. Phys.} \textbf{1926}, \emph{37}, 895.

\bibitem{Gordon} Gordon, W. 
Der Comptoneffekt nach der Schrodingerschen Theorie. 
\emph{Z. Phys.} \textbf{1926}, \emph{40}, 117.

\bibitem{Goldstone} Goldstone, J.; Salam, A.; Weinberg, S. 
Broken Symmetries. \emph{Phys. Rev.} \textbf{1962}, \emph{127}, 965.

\bibitem{Higgs} Higgs, P.W. 
Broken symmetries, massless particles and gauge fields. 
\emph{Phys. Lett.} \textbf{1964}, \emph{12}, 132

\bibitem{Varma} Pekker D.; Varma C.M. 
Amplitude/Higgs Modes in Condensed Matter Physics. 
\emph{Ann. Rev. Cond. Matt. Phys.} \textbf{2015}, \emph{6}, 269.  

\bibitem{Leutwyler} 
{
Leutwyler H. 
Phonons as Goldstone Bosons. \emph{Helv. Phys. Acta} \textbf{1997}, 
\emph{70}, 275. 
}

\bibitem{Brauner} 
{
Brauner T.  
Spontaneous Symmetry Breaking and Nambu–Goldstone Bosons in 
Quantum Many-Body Systems. \emph{Symmetry} \textbf{2010}, 
\emph{2}, 609. 
}
 
\bibitem{Endlich} 
{
Endlich S, Nicolis A., Penco R. 
Ultraviolet completion without symmetry restoration, 
Phys. Rev. D \textbf{2014}, \emph{89}, 065006.
}

\bibitem{Watanabe} 
{
Watanabe H., Murayama H. 
Effective Lagrangian for Nonrelativistic Systems, 
Phys. Rev. X \textbf{2014}, \emph{4}, 031057. 
}


\bibitem{Nielsen} 
{
Nielsen, H.B.; Chadha, S. 
On how to count Goldstone bosons. 
\emph{Nucl. Phys. B} \textbf{1976}, \emph{105}, 445.
}

\bibitem{Sachdev} Sachdev, S. {\it Quantum Phase Transitions};  
Cambridge University Press: Cambridge, UK, 2011. 

\bibitem{Salasnich} Salasnich, L.; Toigo, F. 
Zero-point energy of ultracold atoms. \emph{Phys. Rep.} \textbf{2016}, 
\emph{640}, 1. 

\bibitem{Bogoliubov} Bogoliubov, N.N. On the theory of superfluidity. 
\emph{J. Phys. (USSR)} \textbf{1947}, \emph{11}, 23.

\bibitem{Kapusta2} Kapusta, J.I. 
Bose-Einstein condensation, spontaneous symmetry breaking, 
and gauge theories. \emph{Phys. Rev. D} \textbf{1981}, \emph{24}, 2.

\bibitem{Bernstein} Bernstein, J.; Dodelson, S. 
Relativistic Bose gas. \emph{Phys. Rev. Lett.} \textbf{1991}, \emph{66}, 6.
  
\bibitem{Alford2} Alford, M.G.; Mallavarapu, S.K.;  
Schmitt, A.; and Stetina, S. 
From a complex scalar field to the two-fluid picture of superfluidity. 
\emph{Phys. Rev. D} \textbf{2014}, \emph{89}, 085005.

\bibitem{Cea} Cea, T.; Castellani, C.; Seibold, G.; L. Benfatto, L. 
Nonrelativistic Dynamics of the Amplitude (Higgs) Mode 
in Superconductors. \emph{Phys. Rev. Lett.} \textbf{2015}, 
\emph{115}, 157002. 

\bibitem{Gersch} Gersch, H.; Knollman, G. Quantum Cell Model for Bosons. 
Phys. Rev. \textbf{1963}, \emph{129}, 959.

\bibitem{Hubbard} Hubbard, J. (1963). 
Electron Correlations in Narrow Energy Bands. 
Proc. Royal Soc. of London. \textbf{1963}, \emph{276}, 238.

\bibitem{Dupuis} Sengupta, K.; Dupuis, N. 
Mott insulator to superfluid transition in the Bose-Hubbard 
model: a strong-coupling approach. 
\emph{Phys. Rev. A} \textbf{2005}, \emph{71}, 033629.

\bibitem{Endres} Endres, M.; Fukuhara, T.; Pekker, D.; 
Cheneau, M.; Schaub, P.; Gross, C.; Demler, E.; Kuhr, S.; Bloch, I. 
The Higgs Amplitude Mode at the Two-Dimensional Superfluid-Mott 
Insulator Transition. \emph{Nature} \textbf{2012}, \emph{487}, 454.
 
\end{thebibliography}
\end{document}